\documentclass[aps,prl,reprint,showpacs,byrevtex]{revtex4-2}
\let\oldhat\hat
\renewcommand{\hat}[1]{\oldhat{\mathbf{#1}}}
\usepackage{titlesec}
\usepackage{array}

\usepackage{times,mathptm}% for making `pdf' file
\usepackage[dvips]{graphicx,color}% for inclusion of graphics file(s)

\usepackage{amsmath}
\usepackage{xcolor}

\begin{document}
\title{Stacking-tunable multiferroic states in bilayer ScI$_2$}
\author{Yaxin Pan$^1$, Chongze Wang$^2$, Shuyuan Liu$^{2}$, Fengzhu Ren$^1$, Chang Liu$^1$, Bing Wang$^{1*}$, and Jun-Hyung Cho$^{1,2*}$}
\affiliation{$^1$Joint Center for Theoretical Physics, School of Physics and Electronics, Henan University, Kaifeng 475004, China \\
$^2$Department of Physics and Research Institute for Natural Science, Hanyang University, 222 Wangsimni-ro, Seongdong-Ku, Seoul 04763, Repu\emph{}blic of Korea}
\date{\today}

\begin{abstract}
Two-dimensional (2D) multiferroic materials hold significant promise for advancing the miniaturization and integration of nanodevices. In this study, we demonstrate that 2D bilayer ScI$_2$, which exhibits ferromagnetic (FM) ordering within each layer, enables the tuning of interlayer magnetic coupling, ferroelectricity, and valley polarization through interlayer sliding and rotation. Our first-principles calculations show that the AA stacking configuration induces antiferromagnetic (AFM) interlayer coupling, while a 180$^{\circ}$ rotation of one layer (resulting in the antialigned AA$^{*}$ stacking) leads to FM interlayer coupling. Moreover, the interlayer magnetic coupling can be switched between AFM and FM by translating the stacking configuration: FM in the aligned AB and BA configurations, and AFM in the antialigned AB$^{*}$ and BA$^{*}$ configurations. This switching behavior is driven by variations in superexchange interactions due to orbital hopping between layers. Notably, the aligned stacking exhibits ferroelectricity upon sliding, which is induced by interlayer orbital hybridization and the resulting asymmetric charge redistribution, with maximal ferroelectric behavior occurring at the AB and BA stacking configurations. Additionally, for the AB and BA (AB$^{*}$ and BA$^{*}$) stackings, spontaneous valley polarization emerges from the manipulation of the spin orientation toward the out-of-plane direction. This valley polarization arises due to inversion symmetry breaking, either through ferroelectricity (in the AB and BA stackings) or AFM interlayer coupling (in the AB$^{*}$ and BA$^{*}$ stackings), in combination with spin-orbit coupling. These results highlight the intricate interplay between magnetism, ferroelectricity, and valley polarization in bilayer ScI$_2$, with each property being tunable via stacking configuration. Our findings offer valuable insights into the design of 2D multiferroic devices, opening up possibilities for applications in next-generation spintronic, electronic, and valleytronic technologies.
\end{abstract}
\pacs{}
\maketitle
%\begin{multicols}{2}

%\vspace{0.4cm}
\section{I. INTRODUCTION}
%\vspace{0.4cm}

Two-dimensional (2D) van der Waals (vdW) materials have emerged as promising platforms for next-generation nanodevices due to their structural versatility, reduced dimensionality, and tunable electronic properties~\cite{Graphene-Science2004,Graphene-Rev2009,TMDCs-Rev2017,2D-herer.-Science2016}. The weak interlayer vdW interactions in these materials enable precise control over various physical properties through stacking engineering, making them suitable for a wide range of applications in electronics, magnetism, and optics. Recent studies have shown that structural modifications such as interlayer rotation and sliding can induce novel quantum states by altering symmetry, charge distribution, and electronic correlations~\cite{polytypes,Ultrafast}. Twisted bilayer systems, including graphene~\cite{Nature2019,PRL2019,NP2021,Nature572} and transition metal dichalcogenides~\cite{NM2020,NP2020,Nature637,NatureWSe2}, exhibit moir\'e-induced electronic reconstructions, leading to correlated electronic states, superconductivity, and topological bands. Meanwhile, stacking variations through interlayer sliding drive magnetic phase transitions, as observed in bilayers CrI$_3$~\cite{PRB99,NL18} and CrBr$_3$~\cite{JPCC125,Science366}, where different stacking configurations modify interlayer magnetic coupling. Sliding-induced ferroelectricity has also been reported in bilayer MoS$_2$~\cite{NC13,PRL133}, HgI$_2$ ~\cite{NL24}, and GdI$_2$~\cite{NL2024} where nonpolar monolayers transform into polar bilayers due to symmetry breaking caused by the stacking. Additionally, stacking engineering has proven effective in controlling valley polarization, as demonstrated in ferrovalley materials like YI$_2$~\cite{NL23} and VSiGeP$_4$~\cite{PCCP26}. These discoveries highlight the critical role of stacking order, interlayer sliding, and rotation in tuning exotic quantum states and ferroic properties. By leveraging these structural degrees of freedom, 2D materials offer a versatile platform for designing next-generation electronic, spintronic, and valleytronic devices.

Multiferroic materials, which exhibit coupled ferroic orders such as ferromagnetism (or antiferromagnetism), ferroelectricity, and ferrovalley properties, present great promise for next-generation electronic and spintronic applications. Many existing 2D multiferroic systems are based on heterostructures such as CrI$_3$/In$_2$Se$_3$~\cite{PRBCri3/In2Se3}, Cr$_2$Ge$_2$Te$_6$/In$_2$Se$_3$~\cite{NC10}, LaCl/In$_2$Se$_3$~\cite{NC11}, LaBr$_2$/In$_2$Se$_3$~\cite{APL125}, and EuSn$_2$As$_2$/In$_2$Se$_3$~\cite{PRB110}, but the fabrication complexities of these layered assemblies pose challenges for large-scale integration and practical implementation. A more effective strategy is to achieve intrinsic multiferroicity through stacking engineering within a single material~\cite{PRB99,NL18,JPCC125,Science366}. This approach not only simplifies fabrication but also enables precise and tunable control over the coupling between multiple ferroic orders. By harnessing interlayer sliding and rotational degrees of freedom, it becomes possible to dynamically manipulate magnetic, ferroelectric, and valleytronic properties, unlocking new functionalities in 2D materials. These advancements pave the way for compact, reconfigurable, and multifunctional devices, with significant potential for future nanoscale electronic and quantum technologies. Using a k·p model analysis and first-principles calculations, He et al.~\cite{APLScI2} showed that monolayer ScI$_2$ exhibits both an anomalous valley Hall effect and a valley‑polarized quantum anomalous Hall effect under~4.7\% tensile strain; these valley‑related Hall effects vanish when the strain is increased further, illustrating their simultaneous emergence and disappearance. Recently, Dong and Zhang ~\cite{PCCPScI2} reported both ferroelectricity and ferrovalley properties, with its valley polarization tunable via interlayer sliding symmetry. Here, we choose bilayer ScI$_2$, where each Sc atom carries a 3$d^1$ electron that endows the material with intrinsic magnetism. By exploiting interlayer sliding and rotational degrees of freedom, we demonstrate that magnetic, ferroelectric, and valleytronic properties can be dynamically and coherently manipulated, providing a unified platform for multifunctional 2D quantum materials.

In this study, we use first-principles calculations to show that the electronic and magnetic properties of bilayer ScI$_2$, where each layer exhibits ferromagnetic (FM) ordering, are highly sensitive to stacking configurations, making it a versatile platform for multiferroic behavior. The AA stacking configuration stabilizes antiferromagnetic (AFM) interlayer coupling, while a 180$^{\circ}$ rotation (antialigned AA$^{*}$ stacking) induces FM interlayer coupling. Moreover, lateral translation from the AA (AA$^{*}$) to the AB and BA (AB$^{*}$, BA$^{*}$) stackings enables AFM-FM (FM-AFM) switching through superexchange interactions. In addition to magnetism, stacking modifications also induce ferroelectricity and valley polarization. The aligned stackings exhibit ferroelectricity upon sliding from the AA stacking due to interlayer orbital hybridization and asymmetric charge redistribution, with the strongest response predicted at the AB and BA stackings. However, ferroelectricity is absent in the AA$^{*}$, AB$^{*}$, and BA$^{*}$ stacking configurations. Furthermore, in the AB and BA (AB$^{*}$ and BA$^{*}$) stacking configurations, spontaneous valley polarization arises from the manipulation of spin orientation toward the out-of-plane direction, resulting from inversion symmetry breaking via ferroelectricity or AFM interlayer coupling, in conjunction with spin-orbit coupling (SOC). These findings underscore the intricate interplay between magnetism, ferroelectricity, and valley polarization in bilayer ScI$_2$, highlighting stacking as a crucial control parameter for engineering multifunctional 2D materials in next-generation spintronic, electronic, and valleytronic devices.

%\vspace{0.4cm}
\section{II. CALCULATIONAL METHODS}
%\vspace{0.4cm}

Our first-principles density-functional theory (DFT) calculations were performed using the Vienna \textit{ab initio} simulation package (VASP) with a plane-wave basis \cite{vasp1,vasp2}. The interactions between core and valence electrons were described using the projector-augmented wave (PAW) method \cite{paw}. The exchange-correlation energy was treated within the generalized gradient approximation (GGA) using the Perdew-Burke-Ernzerhof (PBE) functional \cite{pbe}. A kinetic energy cutoff of 500 eV and a total energy convergence criterion of $10^{-8}$ eV were employed. Atomic positions were fully relaxed until the maximum force on each atom was less than 0.01 eV/{\AA}. To prevent interactions between adjacent layers in a periodic slab geometry, a vacuum spacing of 20 {\AA} in the out-of-plane direction was used. An $18 \times 18$ $\mathbf{k}$-point mesh was employed in the 2D Brillouin zone. Van der Waals interactions were treated using the DFT-D3 method \cite{JCP154104,JCC1456}. To describe the localized 3$d$ electrons of Sc, we performed DFT + $U$ calculations using the Dudarev approach \cite{Dudarev}, where the Hubbard Coulomb interaction $U$ was set to 2.5 eV for the Sc 3$d$ orbitals \cite{AM246}. Berry curvature is obtained by computing the $\mathbf{k}$-space derivatives of the Wannier-interpolated Hamiltonian using the WannierTools package~\cite{wanniertool}, based on maximally localized Wannier functions generated by Wannier90~\cite{wannier90}.

%\vspace{0.4cm}

%\vspace{0.4cm}
\section{III. RESULTS and DISCUSSION}
%\vspace{0.4cm}
\noindent \subsection{1. Magnetism}

 We begin by investigating the magnetic properties of monolayer ScI$_2$ in the 1H phase. The 1H phase adopts hexagonal crystal symmetry (space group $P$-$6m2$), where a single triangular Sc layer is sandwiched between two triangular I layers, with the Sc atoms in trigonal-prismatic coordination with the I atoms, as shown in Fig. 1(a). Similar to other 2D monolayers with comparable $d$-valence electron configurations, such as LaBrI, CeI$_2$, and GdI$_2$~\cite{PRB104,PRB105,MHGdI2}, the Sc atoms in monolayer ScI$_2$ possess a 3$d^1$ electronic configuration, which contributes to the material's magnetic properties. Due to the ligand-field splitting of the $d$ levels in trigonal-prismatic coordination, this 3$d^1$ electron predominantly occupies the $d_{z^2}$ orbital, as discussed below. Our spin-polarized calculations for monolayer ScI$_2$ reveal that the FM state is more stable than the AFM state by 13.7 meV per formula unit (f.u.). The FM state has lattice constants of \(a = b = 4.091 \, \text{\AA}\) and is thermodynamically stable, as confirmed by the absence of imaginary phonon modes in the phonon dispersion [see Fig. 2(a)] and the structural integrity shown in {\em ab initio} molecular dynamics simulations [see Fig. 2(b)]. Figure 2(c) shows the calculated band structure and partial density of states (PDOS) of the FM state, where the Sc 3$d_{z^2}$ orbital splits into spin-up and spin-down bands, with a separation of approximately 0.28 eV near the Fermi level $E_F$. This splitting is a characteristic feature of the FM instability and contributes to the formation of a band gap. The I 5$p$ orbitals are fully occupied and primarily located around 3.5~eV below $E_F$ [see Fig.~2(c)], indicating their minimal contribution to the bonding, which is predominantly ionic in character. It is, however, noted that the I 5$p$ states are weakly hybridized with the top of the valence band, which is mainly composed of the Sc 3$d_{z^2}$ orbital~\cite{orbit}. Furthermore, monolayer ScI$_2$ exhibits an in-plane magnetic easy axis, indicating a preferred alignment of magnetic moments within the $xy$-plane [see Supplemental Fig. S2(a)~\cite{SM}]. Additionally, Monte Carlo simulations based on the Heisenberg model estimate the Curie temperature for the FM order to be approximately 85 K [see Supplemental Fig. S2(b)~\cite{SM}].

\begin{figure}[h!t]
\includegraphics[width=8.5cm]{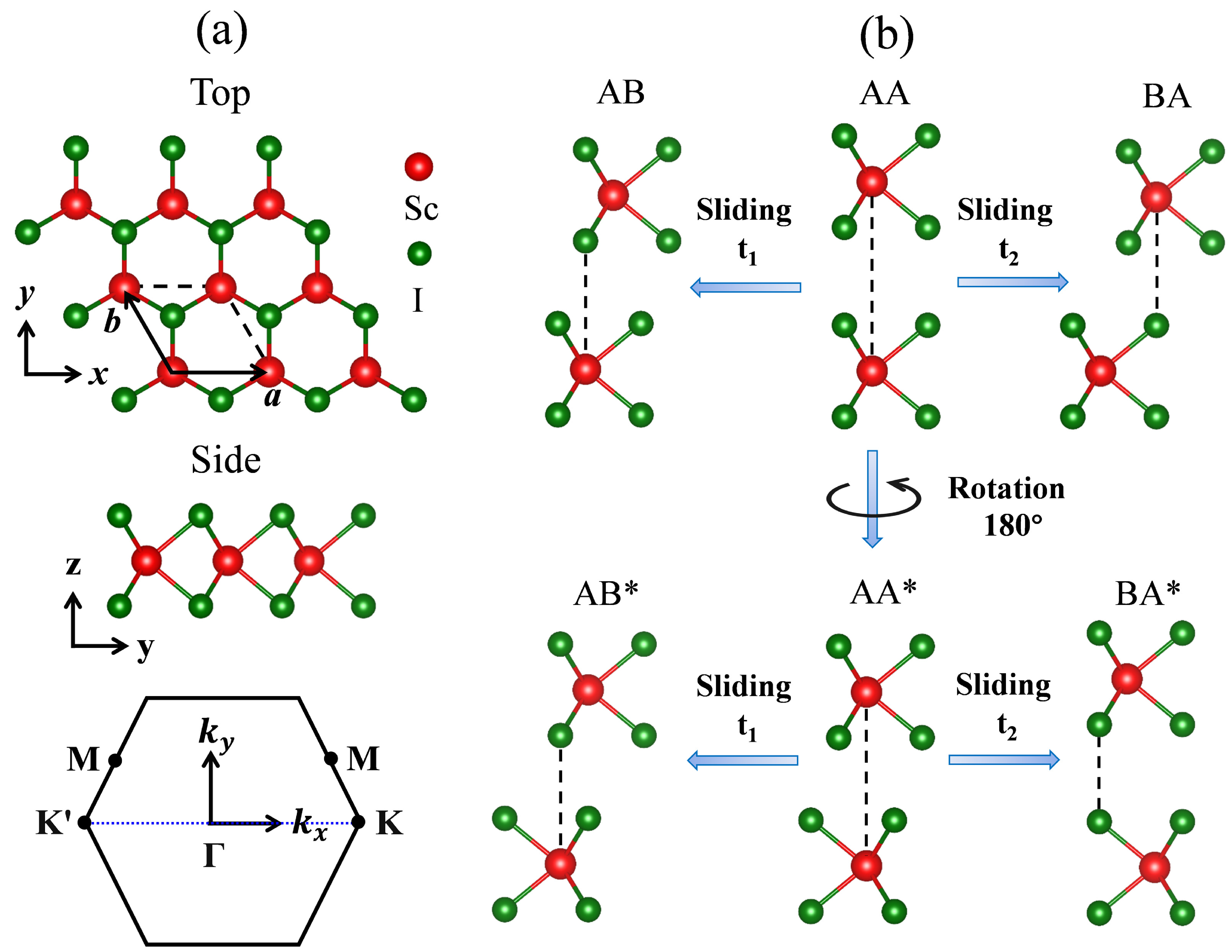}
\caption{Optimized structures of (a) monolayer and (b) bilayer ScI$_2$. For the bilayer ScI$_2$, six stacking configurations are displayed: three aligned stacking configurations (AA, AB, and BA) and three antialigned stacking configurations (AA$^{*}$, AB$^{*}$, and BA$^{*}$). In (a), the lattice parameters \(a\) and \(b\) represent the primitive unit cell, along with its corresponding Brillouin zone. In (b), \( \mathbf{t}_1 = \left( -\frac{1}{3}, -\frac{2}{3} \right) \) and \( \mathbf{t}_2 = \left( \frac{1}{3}, \frac{2}{3} \right) \) denote fractional in-plane translations of the upper layer along the \(y\)-axis.}
\label{figure:1}
\end{figure}

\begin{figure}[h!t]
\includegraphics[width=8.5cm]{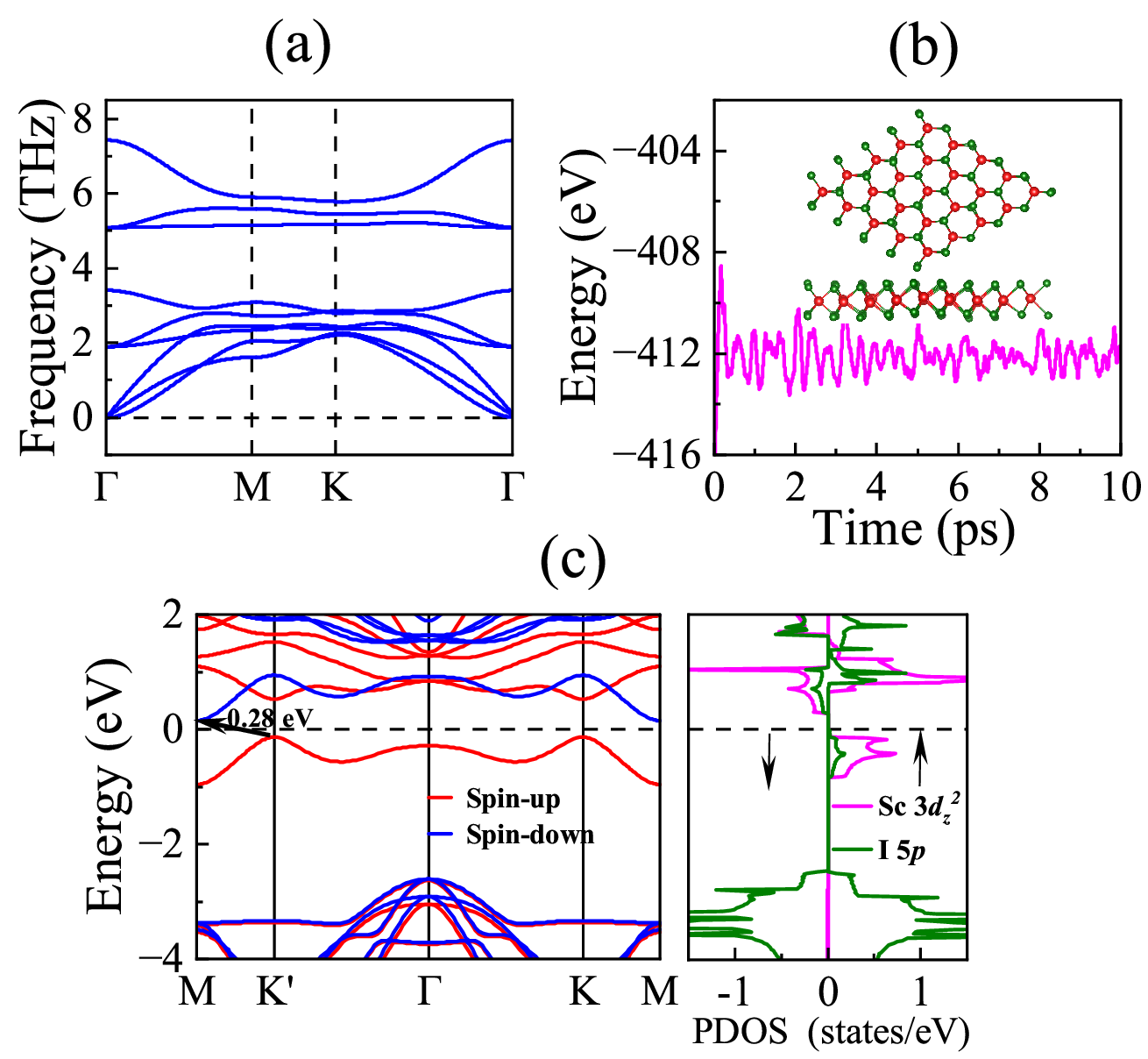}
\caption{Calculated (a) phonon dispersion, (b) total energy variation from {\em ab initio} molecular dynamics simulations, and (c) spin-polarized band structure of monolayer ScI$_2$. The inset in (b) shows the structure at the end of the simulation. The PDOS for the Sc 3$d_{z^2}$ and I 5$p$ orbitals is also provided in (c), while that for other orbitals is shown in Supplemental Fig. S1~\cite{SM}.}
\label{figure:2}
\end{figure}

Next, we investigate the magnetic ground states of the six stacking configurations in bilayer ScI$_2$. Figure 1(b) shows the aligned stacking configurations (AA, AB, and BA) as well as the antialigned stacking configurations (AA$^{*}$, AB$^{*}$, and BA$^{*}$). The AA stacking in bilayer ScI$_2$ is formed by directly placing one layer on top of the other, preserving mirror symmetry \( M_z \) in the \( xy \)-plane. The AB and BA stackings are derived from the AA stacking by laterally translating the upper layer by \( \mathbf{t_1} = \left(-\frac{1}{3}, -\frac{2}{3} \right) \) and \( \mathbf{t_2} = \left( \frac{1}{3}, \frac{2}{3} \right) \), respectively. Note that the BA stacking configuration can be obtained by a 180$^{\circ}$ rotation of the AB stacking configuration about the $y$-axis. The AA$^{*}$ stacking is obtained by rotating the lower layer in the AA stacking by 180$^{\circ}$ around the \( xy \)-plane, preserving inversion ($P$) symmetry. Similarly, the AB$^{*}$ and BA$^{*}$ stacking configurations are derived from the AA$^{*}$ stacking configuration by applying the same lateral translations \(\mathbf{t_1}\) and \(\mathbf{t_2} \), respectively [see Fig. 1(b)]. Figures 3(a) and 3(b) present the energy profiles of the interlayer FM and AFM coupling states for the aligned and antialigned stacking configurations, respectively, as a function of interlayer sliding. As shown in Fig. 3(a), the AB and BA stackings exhibit the same lowest energy, which is $34.24$ meV/f.u. lower than the AA stacking. The transition from the AB to BA stacking requires overcoming an energy barrier of $8.87$ meV/f.u. We find that the AA stacking favors the AFM state over the FM state by 0.50 meV/f.u., while the AB and BA stackings favor the FM state over the AFM state by 0.33 meV/f.u. Conversely, in the antialigned stackings, the AA$^{*}$ stacking favors the FM state over the AFM state by $0.11$ meV/f.u., whereas the AB$^{*}$ and BA$^{*}$ stackings favor the AFM state over the FM state by $0.53$ and $0.29$ meV/f.u., respectively [see Fig. 3(b)]. Notably, the AA$^{*}$ and AB$^{*}$ stackings are more stable than BA$^{*}$ by 33.39 and 34.80 meV/f.u., respectively~\cite{distance}. These results indicate that the energy difference between the FM and AFM states varies continuously across different stacking configurations, demonstrating that the magnetic ground state transitions between FM and AFM as the stacking order changes [see Figs. 3(a) and 3(b)].

\begin{figure}[h!t]
\includegraphics[width=8.5cm]{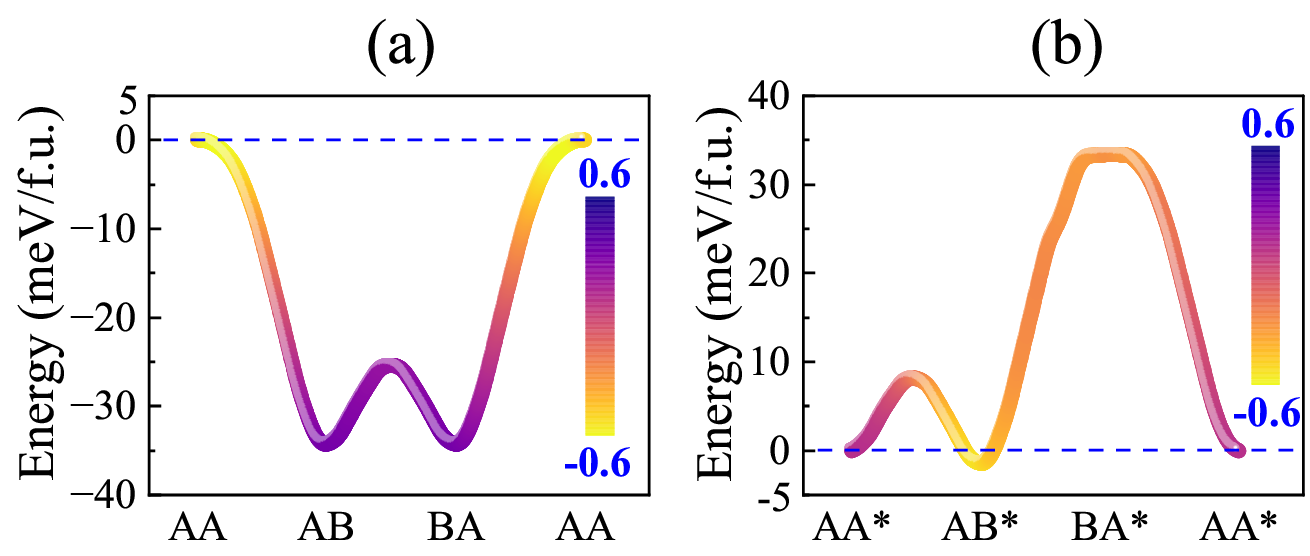}
\caption{Calculated energy profiles of the interlayer FM and AFM coupling states for the (a) aligned and (b) antialigned stacking configurations as a function of interlayer sliding. The zero energy in (a) and (b) is referenced to the AA and AA$^{*}$ stacking configurations, respectively. The color scale represents the energy difference between the interlayer FM and AFM coupling states, where a positive (negative) value indicates a preference for the FM (AFM) state, with the unit of meV/f.u.}
\label{figure:3}
\end{figure}

To elucidate the switching behavior between FM and AFM interlayer couplings across different stacking configurations, we estimate the intralayer (\( J_{1} \)), first interlayer (\( J_{\text{inter-1}} \)), and second interlayer (\( J_{\text{inter-2}} \)) Sc–Sc exchange interactions. Here, \( J_{\text{inter-1}} \) and \( J_{\text{inter-2}} \) correspond to the nearest-neighbor (NN) and next-nearest-neighbor (NNN) interactions between the two layers, respectively. The exchange parameters \( J_{1} \), \( J_{\text{inter-1}} \), and \( J_{\text{inter-2}} \), derived from the Heisenberg spin Hamiltonian, are obtained through total energy calculations based on DFT (see their derivations in the Supplemental Material~\cite{SM}), and the results for each stacking configuration are summarized in Table I. For all stacking configurations, \( J_{1} \) is negative and has a magnitude larger than 67~meV (i.e., \( J_{1} < -67 \)~meV), indicating a strong preference for FM ordering within each layer. However, the interlayer exchange interactions vary significantly depending on the stacking configurations. In the AA stacking, we obtain \( J_{\text{inter-1}} = 2.564 \)~meV and \( J_{\text{inter-2}} = -0.066 \)~meV. Since the AA stacking has only one NN and six NNNs [see Fig. 4(a)], the total interlayer exchange interaction is 2.168 meV, stabilizing an AFM ground state. In contrast, for the AB (or BA) stacking, where \( J_{\text{inter-1}} = -0.249 \) meV with three NNs and \( J_{\text{inter-2}} = -0.186 \)~meV with three NNNs [see Fig. 4(b)], the total interlayer exchange interaction sums to \( -1.305 \) meV, favoring an FM ground state. A similar FM preference is predicted in the AA$^{*}$ stacking, with \( J_{\text{inter-1}} = 0.029 \)~meV (one NN) and \( J_{\text{inter-2}} = -0.078 \)~meV (six NNNs), as shown in Supplemental Fig. S4(a), resulting in a total interlayer interaction of \( -0.439 \) meV. Meanwhile, in the AB$^{*}$ (BA$^{*}$) stacking, the total interlayer exchange interaction is positive, with a value of 2.379 (1.299) meV, where \( J_{\text{inter-1}} = -0.009 \) (0.389) meV with three NNs and \( J_{\text{inter-2}} = 0.802 \) (0.044) meV with three NNNs [see Supplemental Fig. S4(b)~\cite{SM}], stabilizing an AFM ground state. These results highlight the critical role of stacking configuration in determining the interlayer magnetic coupling in bilayer ScI$_2$, revealing the intricate interplay between stacking structures and magnetic interactions.

\begin{figure}[h!t]
\includegraphics[width=8.5cm]{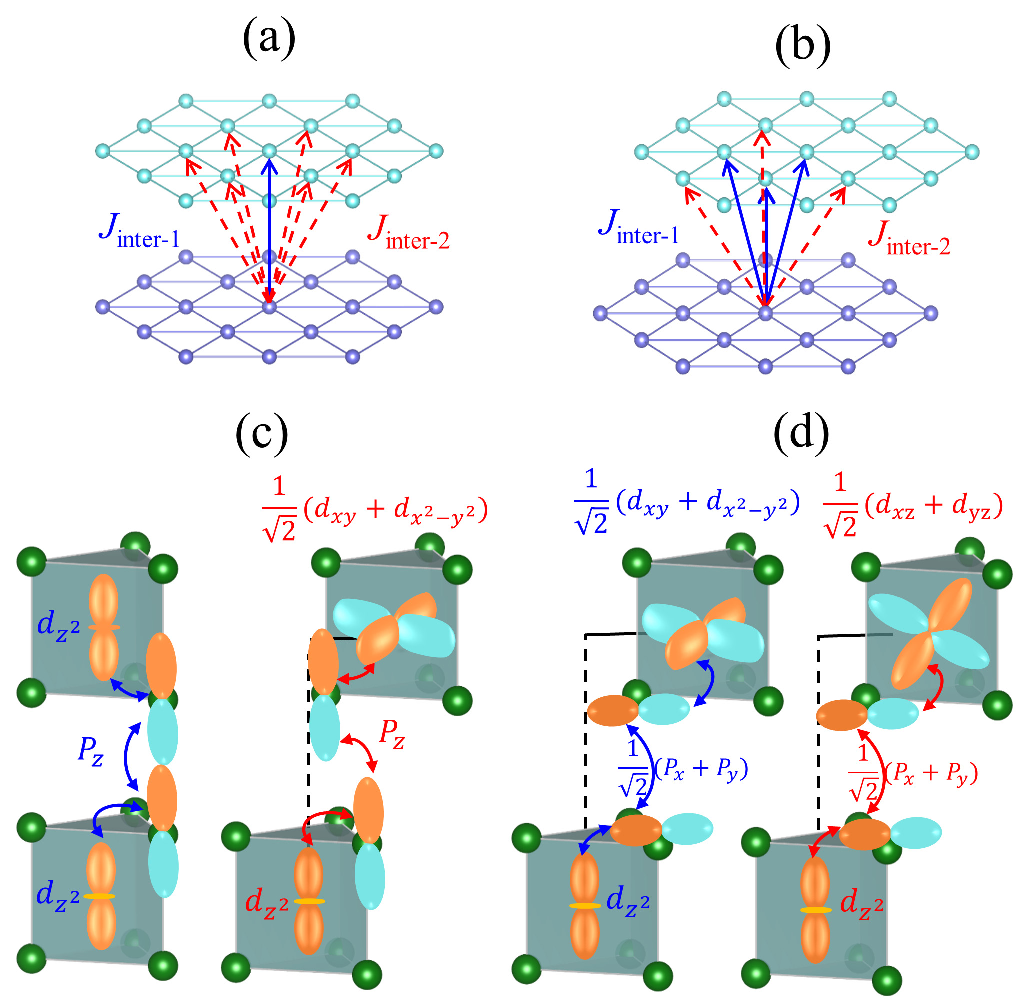}
\caption{The NN and NNN Sc atoms for the (a) AA and (b) AB stacking configurations. The solid (dashed) lines represent the connections between NN (NNN) Sc atoms. The schematic diagrams of the orbitals participating in the supersuperexchange mechanism for the AA and AB stacking configurations are displayed in (c) and (d), respectively.}
\label{figure:4}
\end{figure}

\begin{table}[ht]
\caption{Calculated intralayer (\( J_{1} \)), first interlayer (\( J_{\text{inter-1}} \)), and second interlayer (\( J_{\text{inter-2}} \)) Sc–Sc exchange parameters for the six stacking configurations of bilayer ScI$_2$, along with the corresponding magnetic moments of the Sc atoms.}
\begin{ruledtabular}
%\scriptsize
\begin{tabular}{lrrrr}
         &   \( J_{1} \) (meV)  &   \( J_{\text{inter-1}} \) (meV) &  \( J_{\text{inter-2}} \) (meV) & $m$ (${\mu}_B$) \\ \hline
AA       &  $-$67.682  &    2.564    &  $-$0.066  & 0.584 \\
AB (BA)  &  $-$67.990  & $-$0.249   &  $-$0.186  & 0.585 \\
AA$^{*}$ &  $-$67.900  &    0.029   &  $-$0.078  & 0.585 \\
AB$^{*}$  &  $-$67.716  & $-$0.009   &     0.802  & 0.583 \\
BA$^{*}$  &  $-$67.487  &    0.389  &     0.044  & 0.584 \\
\end{tabular}
\end{ruledtabular}
%\end{ruledtabular}
\end{table}

The interlayer Sc–Sc exchange interactions in bilayer ScI$_2$ are primarily governed by a superexchange mechanism involving Sc 3\(d\) orbitals, mediated by hybridization with I 5\(p\) orbitals. Consequently, the stacking-dependent magnetism arises from the alignment and hybridization of the Sc 3\(d\) and I 5\(p\) orbitals, which determine the exchange pathways and hopping processes responsible for AFM or FM interlayer coupling~\cite{NL18}. In the AA stacking configuration, the Sc atoms in the upper and lower layers are vertically aligned, forming an exchange pathway for \(J_{\text{inter-1}}\) between Sc \(d_{z^2}\) orbitals via I \(p_z\) orbitals [see Fig. 4(c)]. This vertical alignment of I atoms enables strong orbital overlap between the Sc \( d_{z^2} \) and I \( p_z \) orbitals, establishing an efficient interlayer superexchange channel. In this geometry, the \( d_{z^2} \)-\( p_z \)-\( p_z \)-\( d_{z^2} \) hopping process facilitates virtual electron excitations that favor AFM coupling. The AFM preference arises because, in the FM configuration, symmetry constraints suppress the FM exchange interaction between identical \(d_{z^2}\) orbitals due to the Pauli exclusion principle. In contrast, \( J_{\text{inter-2}} \) occurs between laterally displaced Sc atoms in adjacent layers. This geometric offset facilitates virtual excitations between Sc \( d_{z^2} \) and \( \frac{1}{\sqrt{2}}(d_{xy} + d_{x^2-y^2}) \) orbitals via I \( p_z \) orbitals [see Fig.~4(c)]. These interactions favor FM coupling due to the orthogonality of the involved \( d \) orbitals, along with the influence of Hund’s rule coupling mediated by the I \( p_z \) orbitals. Hund’s rule promotes parallel spin alignment during the virtual hopping process, thereby favoring FM alignment. However, since \(J_{\text{inter-1}}\) is much stronger than the cumulative effect of the six \(J_{\text{inter-2}}\) interactions (see Table I), the dominant nearest-neighbor AFM coupling mediated by \(J_{\text{inter-1}}\) ultimately stabilizes the AFM ground state in AA-stacked ScI$_2$. For the AB (or BA) stacking configuration, where the upper layer is laterally translated by \( \mathbf{t_1} \) (\( \mathbf{t_2} \)), the \(J_{\text{inter-1}}\) exchange pathway exists between Sc \(d_{z^2}\) and \( \frac{1}{\sqrt{2}} (d_{xy} + d_{x^2-y^2}) \) orbitals via I \( \frac{1}{\sqrt{2}} (p_x + p_y) \) orbitals [see Fig. 4(d)], favoring FM coupling. This arises due to the lateral displacement of the Sc atoms, which enables hybridization between Sc \(d\) orbitals and I \(p\) orbitals oriented in the \(x\) and \(y\) directions. The orthogonality between the involved \(d\)-orbitals and the geometry of the hopping process contribute to the preference for FM coupling. Additionally, the \(J_{\text{inter-2}}\) exchange pathway emerges between Sc \(d_{z^2}\) and \( \frac{1}{\sqrt{2}} (d_{xz} + d_{yz}) \) orbitals via I \( \frac{1}{\sqrt{2}} (p_x + p_y) \) orbitals [see Fig. 4(d)], also promoting FM coupling. The combined effects of \(J_{\text{inter-1}}\) and \(J_{\text{inter-2}}\) interactions stabilize the FM ground state in AB- or BA-stacked ScI$_2$, as shown in Fig. 3(a). For the antialigned AA$^{*}$, AB$^{*}$, and BA$^{*}$ stacking configurations, the magnetic superexchange interactions switch between FM and AFM interlayer couplings compared to the aligned AA, AB, and BA stacking configurations, primarily due to modifications in the \(J_{\text{inter-1}}\) and \(J_{\text{inter-2}}\) parameters (see Table I). The 180$^{\circ}$ rotations of the lower layer in the antialigned stacking configurations alter the relative orientations of the Sc 3\(d\) and I 5\(p\) orbitals, which modify orbital hybridizations and hopping pathways, thereby affecting the system's symmetry. These changes in symmetry influence the exchange interactions, which can either enhance or suppress FM or AFM coupling by altering the orbital overlap and the relative alignment of spins across layers. This highlights that the interlayer magnetic interactions are governed by orbital hybridizations and hopping processes between the Sc 3\(d\) and I 5\(p\) orbitals, which determine the dominant exchange pathways and the spin-dependent selection rules that govern the magnetic coupling.

\noindent \subsection{2. Ferroelectricity}

It is well known that in bilayer MoS$_2$, breaking the mirror symmetry \(M_z\) in the D\(_{3h}\) point group induces out-of-plane electric polarization~\cite{NC13}. Similarly, in bilayer ScI$_2$, the AB and BA stacking configurations break both \(M_z\) and \(P\) symmetries, leading to a net polarization. In contrast, the AA stacking configuration breaks $P$ symmetry but preserves \(M_z\) symmetry, preventing the formation of a net polarization, as the polarization induced by $P$ symmetry breaking is canceled out by the \(M_z\) symmetry. As shown in Fig. 5(a), unlike the AA stacking configuration, the AB (BA) stacking configuration exhibits a planar-averaged electrostatic potential difference, with a positive (negative) value of \(\Delta \phi = 29.78\) ($-$29.78) meV between the vacuum levels of the upper and lower layers. Using the Berry phase method~\cite{Vanderbilt}, we calculate the electric polarization of the AB (BA) stacking configuration to be \( 0.18 \times 10^{-12} \)~C/m along the \( +z \) (\( -z \)) axis. For comparison, the electric polarizations in other vdW bilayer ferroelectrics such as NiI$_2$~\cite{NiI2}, YI$_2$~\cite{NL23}, and GdI$_2$~\cite{NL2024} have been reported to be approximately \( 0.44 \times 10^{-12} \), \( 2.6 \times 10^{-12} \), and \( 3.68 \times 10^{-12} \) C/m, respectively. Note that the AB and BA configurations can be interchanged by a 180$^{\circ}$ rotation about the $y$-axis, resulting in a reversed polarization when the stacking is flipped. The underlying cause of these potential differences and the spontaneous polarization along the out-of-plane direction is the simultaneous breaking of both \(M_z\) and $P$ symmetries, accompanied by a redistribution of electronic charge density across the layers. As shown in Fig. 5(b), the charge density difference, defined as \(\Delta \rho = \rho_{\text{bilayer}} - (\rho_{\text{upper-monolayer}} + \rho_{\text{lower-monolayer}})\), reveals distinct patterns between the AA and AB (or BA) stacking configurations. In the AA stacking, \(\Delta \rho\) and its planar average along the \(z\)-axis are symmetric about \(z = 0\), while in the AB and BA stackings, they are asymmetric, resulting in electric polarization along the out-of-plane direction. Since the AB and BA stackings can be interconverted through interlayer sliding, this confirms the presence of sliding ferroelectricity in bilayer ScI$_2$.

\begin{figure}[h!t]
\includegraphics[width=8.5cm]{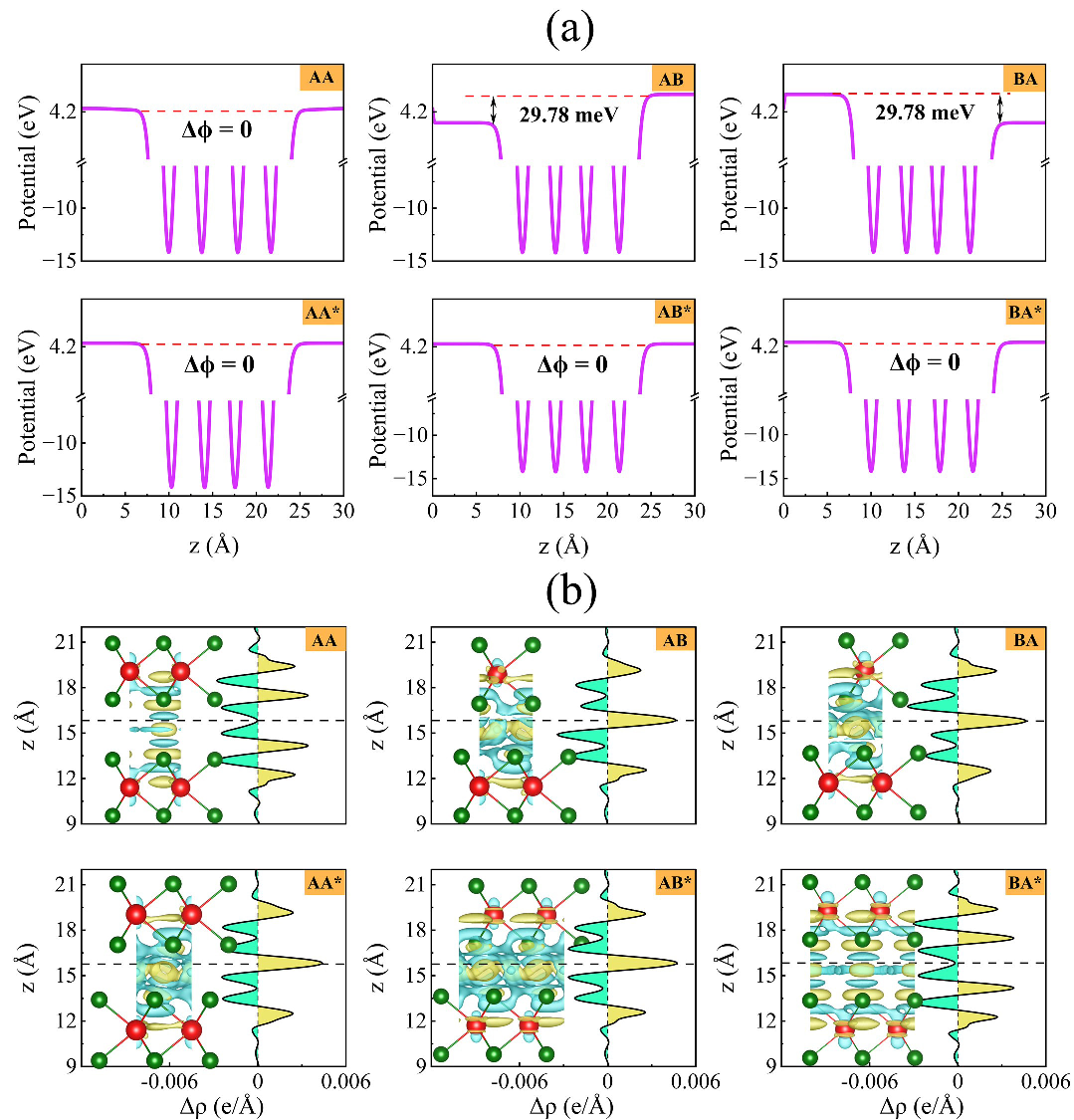}
\caption{(a) Calculated planar averages of electrostatic potentials along the $z$ axis for the aligned and antialigned stacking configurations. The charge density differences ${\Delta}{\rho}$ and their planar averages along the $z$ axis for the aligned and antialigned stacking configurations are displayed in (b). The isosurface value in ${\Delta}{\rho}$ is set to 0.6${\times}$10$^{-4}$ $e$/{\AA}$^3$.}
\label{figure:5}
\end{figure}

For the AA$^{*}$, AB$^{*}$, and BA$^{*}$ stacking configurations, the planar-averaged electrostatic potential profiles along the \( z \)-axis exhibit \(\Delta \phi = 0\) [see Fig.~5(a)], indicating the absence of spontaneous polarization due to the restoration of either $P$ symmetry or combined $P$ and time-reversal ($T$) symmetries. As shown in Fig.~5(b), \(\Delta \rho\) and its planar average along the \( z \)-axis are symmetric about \( z = 0 \), ensuring that any interlayer-induced dipole moments cancel out. Therefore, in these antialigned configurations, symmetry strictly prohibits the emergence of macroscopic electric polarization.

\noindent \subsection{3. Valley polarization}

To investigate the emergence of valley polarization in bilayer ScI$_2$, we calculate the band structures for the aligned and antialigned stacking configurations, considering both cases without and with SOC. Figures~6(a)–6(f) show the calculated band structures for the AA, AB, BA, AA$^{*}$, AB$^{*}$, and BA$^{*}$ configurations, respectively. In the AA stacking configuration, which favors AFM interlayer coupling, the combined \( C_{2y} \) and \( T \) symmetries protect the degeneracy of spin-up and spin-down states along the \( \Gamma \)-\( K \) and \( \Gamma \)-\( K' \) lines in the absence of SOC [see Fig.~6(a)]. This degeneracy arises from the antiunitary nature of the \( C_{2y}T \) operator, satisfying \( (C_{2y}T)^2 = -1 \), which leads to a sign change upon two successive operations. Furthermore, the \( M_z T \) symmetry guarantees the degeneracy between the states at \( \mathbf{k} \) along the \( \Gamma \)-\( K \) line and \( -\mathbf{k} \) along the \( \Gamma \)-\( K' \) line [see the upper panel in Fig.~6(a) and the symmetry analysis in the Supplemental Material~\cite{SM}]. When SOC is included, the band degeneracy remains preserved for in-plane spin orientations due to the protection provided by \( M_z \) symmetry [see Supplemental Fig.~S5(a)~\cite{SM}]. In contrast, for out-of-plane spin orientations, the degeneracy is lifted [see the lower panel in Fig.~6(a)] because the \( M_z \) symmetry is broken. Notably, the small magnetic anisotropy energy of approximately ${\sim}$0.399~meV/f.u. [see Supplemental Fig.~S2(a)] suggests that  external perturbations, such as spin-orbit torque induced by an applied magnetic field or current, can manipulate the spin orientations by overcoming the anisotropy energy barrier~\cite{Valley}. For the out-of-plane spin orientation, no net valley polarization appears in the spin-summed bands because the energy of the spin-up state at the \( K \) point is equal to that of the spin-down state at the \( K' \) point. However, when considering spin-resolved contributions, the spin-up and spin-down bands individually exhibit opposite valley polarizations, each with a magnitude of 99.7~meV. This spin-dependent valley polarization has the contrasting Berry curvatures, \( \Omega_z \), at the \( K \) and \( K' \) points, as illustrated in Fig.~7(a). The opposite signs of \( \Omega_z \) at \( K \) and \( K' \) imply that carriers from different valleys experience opposite anomalous velocities under an applied electric field, potentially enabling spin and valley Hall effects. The spin-dependent band splitting predicted here is reminiscent of Rashba-type splitting in the sense that it involves spin-momentum locking, although its origin is different due to the AFM order and broken \( M_z \) symmetry. The magnitude of the splitting depends on the \( \mathbf{k} \)-values, with the \( M_z T \) symmetry ensuring that the spin-up (spin-down) states at a \( \mathbf{k} \) point along the \( \Gamma \)-\( K \) line are degenerate with the spin-down (spin-up) states at the corresponding \( -\mathbf{k} \) point along the \( \Gamma \)-\( K' \) line. This symmetry constraint leads to the spin-dependent valley polarization structure at the \( K \) and \( K' \) points.

\begin{figure}[h!t]
\includegraphics[width=8.5cm]{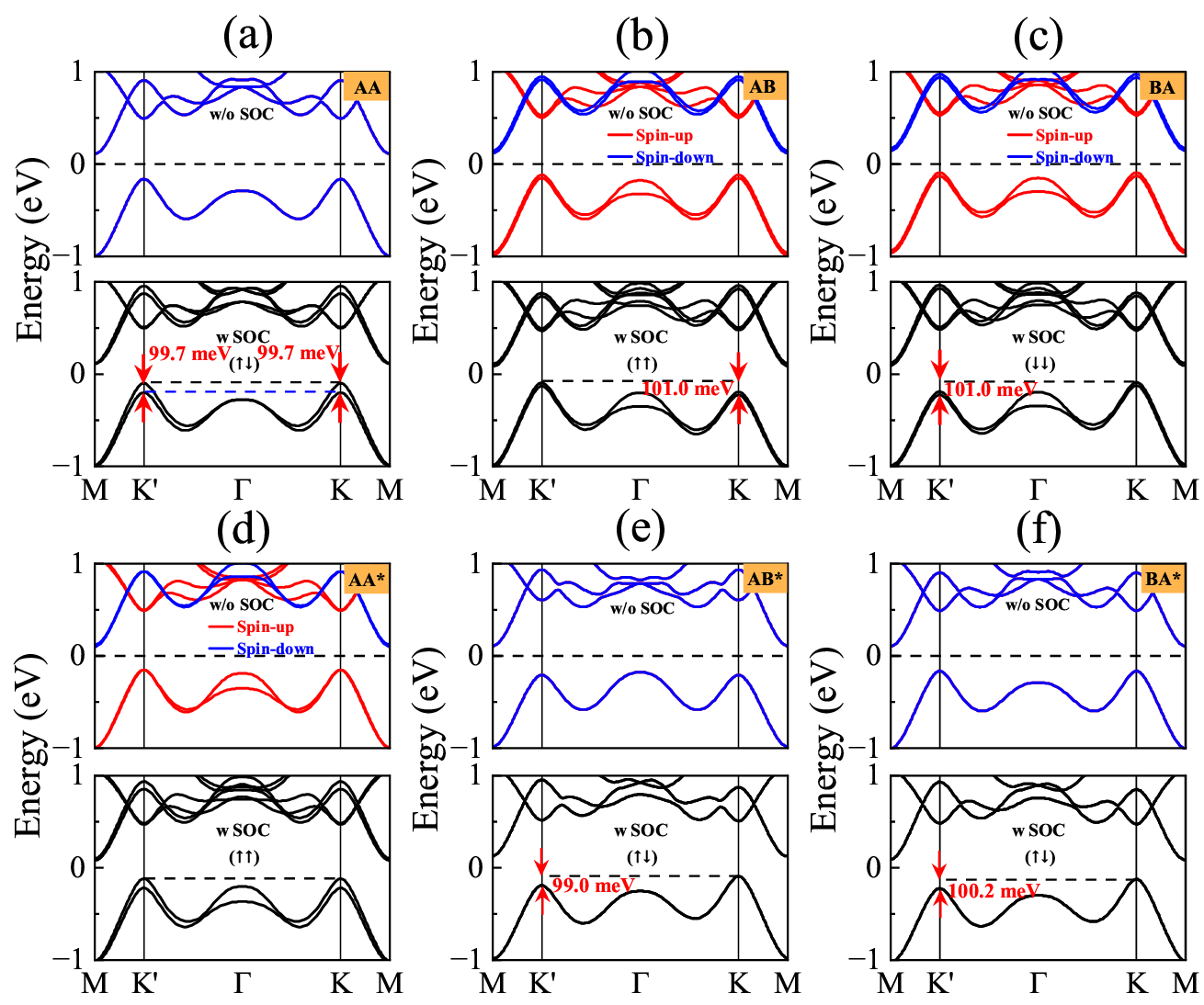}
\caption{Electronic band structures of the (a) AA, (b) AB, (c) BA, (d) AA$^{*}$, (e) AB$^{*}$, and (f) BA$^{*}$ stacking configurations, calculated both without and with SOC. The results without SOC (upper panels) are shown as red and blue lines, representing spin-up and spin-down states, respectively. The results with SOC (lower panels) correspond to out-of-plane spin orientations. Note that the AA, AB$^{*}$, and BA$^{*}$ stacking configurations exhibit degenerate spin-up and spin-down bands in the absence of SOC.}
\label{figure:6}
\end{figure}

For the AB and BA stacking configurations, which exhibit ferroelectricity and FM interlayer coupling, the two topmost valence bands without SOC are spin-polarized as spin-up states [see Figs.~6(b) and 6(c)], originating from the respective layers. The splitting of these bands is induced by charge redistribution due to interlayer hybridization. The mirror symmetry \( M_x \) in the \( yz \)-plane ensures the degeneracy between the spin-up states at the \( \mathbf{k} \) point along the \( \Gamma \)-\( K \) line and the \( -\mathbf{k} \) point along the \( \Gamma \)-\( K' \) line [see Figs.~6(b) and 6(c)], thereby preventing valley polarization. However, with the inclusion of SOC, valley polarization emerges due to the broken \( P \) symmetry: it remains negligible for the in-plane spin orientation [see Supplemental Figs. S5(b) and S5(c)~\cite{SM}] but becomes pronounced for the out-of-plane spin orientation. In the latter case, the valley splitting of the topmost valence band at the \( K \) and \( K' \) points amounts to approximately 101.0~meV for the AB stacking configuration [Fig.~6(b)], and the sign of the valley polarization is reversed in the BA stacking configuration [Fig.~6(c)]. This reversal results from the 180$^{\circ}$ rotational transformation between the AB and BA stacking configurations about the \( y \)-axis, accompanied by their opposite Berry curvatures at the \( K \) and \( K' \) points [see Figs.~7(b) and 7(c)]. Under \( C_{2y} \) symmetry, the \( \Gamma \)-\( K \) and \( \Gamma \)-\( K' \) lines are interchanged, leading to the reversal of valley polarization between the AB and BA configurations.

As shown in Fig.~6(d), the AA$^{*}$ stacking configuration, characterized by FM interlayer coupling, exhibits two topmost valence bands composed of spin-up states. Despite the preservation of \( P \) symmetry, the FM interlayer coupling leads to the formation of bonding and antibonding bands. Both without and with SOC, the \( P \) symmetry guarantees the degeneracy of spin-up states at the \( \mathbf{k} \) point along the \( \Gamma \)-\( K \) line and the \( -\mathbf{k} \) point along the \( \Gamma \)-\( K' \) line [see Figs.~6(d) and Supplemental Fig.~S5(d)], thereby preventing valley polarization at the \( K \) and \( K' \) points. In contrast, the AB$^{*}$ and BA$^{*}$ stacking configurations, exhibiting AFM interlayer coupling, preserve the combined \( PT \) symmetry. This symmetry ensures the degeneracy of spin-up and spin-down bands at every \( \mathbf{k} \) point [see Figs.~6(e) and 6(f)], owing to the antiunitary nature of \( (PT)^2 = -1 \). In the absence of SOC, the \( M_x \) mirror symmetry prevents valley splitting, resulting in no valley polarization. However, when SOC is included, valley polarization remains negligible for the in-plane spin orientation [see Supplemental Figs.~S5(e) and S5(f)] but becomes significant for the out-of-plane spin orientation [Figs. 6(e) and 6(f)]. The corresponding valley polarizations reach values of 99.0 and 100.2~meV for the AB$^{*}$ and BA$^{*}$ configurations, respectively. Thus, the valley polarization predicted in the AB$^{*}$ and BA$^{*}$ stacking configurations arises from the combined effects of \( P \) symmetry breaking, induced by the AFM interlayer coupling, and the contribution of SOC.

\begin{figure}[h!t]
\includegraphics[width=8.5cm]{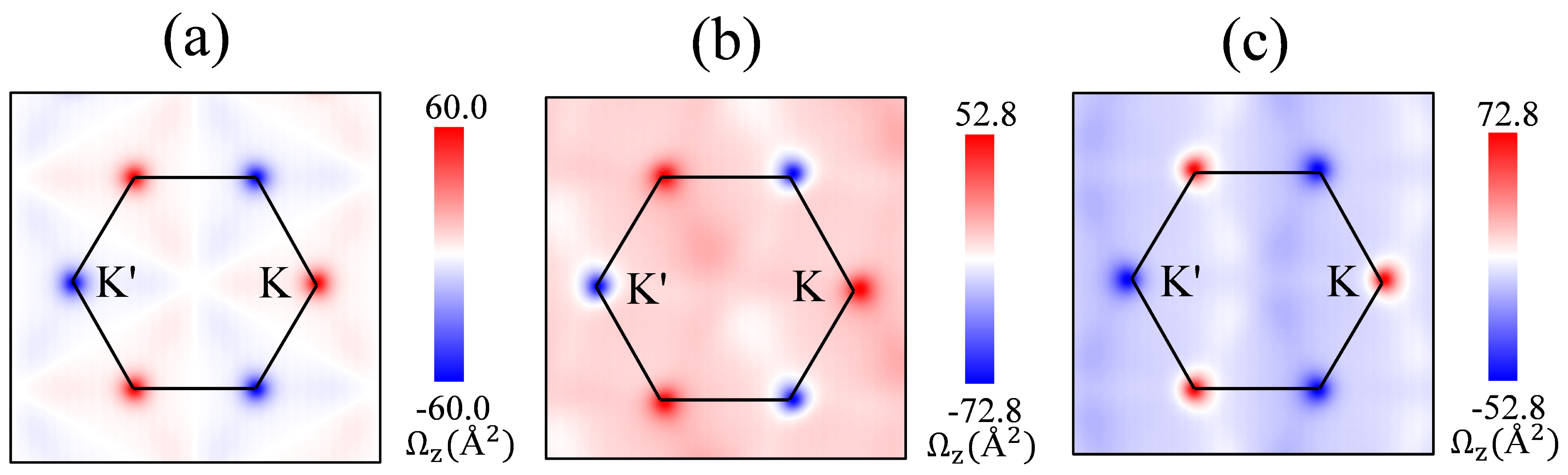}
\caption{Berry curvatures of the topmost valence states for the (a) AA, (b) AB, and (c) BA stacking configurations.}
\label{figure:7}
\end{figure}

%\vspace{0.4cm}
\section{IV. SUMMARY}
%\vspace{0.4cm}

Our first-principles calculations have demonstrated that bilayer ScI$_2$ exhibits a complex interplay of electronic, magnetic, and ferroelectric properties, all of which are highly sensitive to the stacking configurations. Specifically, the AFM interlayer coupling predicted in the AA stacking configuration transitions to the FM interlayer coupling in the AA$^{*}$ configuration via a 180$^{\circ}$ rotation. Furthermore, lateral translations from AA (AA$^{*}$) to AB/BA (AB$^{*}$/BA$^{*}$) lead to AFM-FM (FM-AFM) switching. These changes in the interlayer superexchange interactions are attributed to variations in orbital hybridizations and hoppings, which depend on the stacking configuration. These stacking-dependent transitions were systematically analyzed using the Heisenberg spin Hamiltonian by extracting interlayer exchange parameters, and were found to originate from interlayer superexchange interactions, mediated by orbital hybridization and hopping processes between Sc 3$d$ and  I 5$p$ orbitals. Additionally, the stacking modifications influence both ferroelectricity and valley polarization. The AB and BA stacking configurations exhibit ferroelectricity due to inversion symmetry breaking, which induces charge redistribution across the layers, while valley polarization arises from the combined effects of inversion symmetry breaking and SOC. These findings highlight the versatility of bilayer ScI$_2$ as a multiferroic material, offering opportunities for spintronic, electronic, and valleytronic applications, where stacking configurations play a crucial role in simultaneously tuning multiple functionalities. Our findings may be extended to other 2D vdW bilayers exhibiting similar structural symmetries and electronic configurations, particularly those based on the 1H monolayer phase with partially filled d orbitals, where interlayer sliding or rotation can similarly couple multiple ferroic orders.

\vspace{0.4cm}

\noindent {\bf Acknowledgements.}
This work was supported by the the Natural Science Foundation of Henan (No. 252300421216), the talent Introduction Project in Henan Province (HNGD2025008), the International Cooperation Project of Science and Technology of Henan Province (No.242102520029), and the Foundation of Henan Educational Committee (No.25A140003). J.H.C acknowledges the support from the National Research Foundation of Korea (NRF) grant funded by the Korean Government (Grant No. RS202300218998). The calculations were performed by the KISTI Supercomputing Center through the Strategic Support Program (Program No. KSC-2024-CRE-0055) for the supercomputing application research.

                  %%%%%  REFERENCES  %%%%%
\noindent Y. P. and C. W. contributed equally to this work. \\
\noindent $^{*}$ Corresponding authors: wb@henu.edu.cn and cho@henu.edu.cn


\begin{thebibliography}{99}
\bibitem{orbit} We note that although the Sc 3$d_{z^2}$ orbital has the largest contribution near the valence band edge, strong hybridization with Sc $d_{xy/x^{2}-y^{2}}$ and I $p$ orbitals [see Figs. 2(c) and S1] introduces non-zero angular momentum components, enabling SOC to induce finite valley splitting between the \( K \) and \( K' \) points, as shown in Fig. 6.
\bibitem{Valley} We note that applying an external magnetic field is a well-established experimental technique for tuning magnetic and electronic properties in 2D materials, and has been successfully employed to control valley splitting in related systems ~\cite{Naturenano,SA}. Therefore, although the ground state favors in-plane spin orientation, the valley splitting predicted for out-of-plane spin alignment remains experimentally accessible and may provide valuable guidance for future studies.
\end{thebibliography}
\end{document}